\documentstyle{aipproc}

\begin{document}
\title{The Redshift of GRB 970508}

\author{Daniel E. Reichart}
\address{{Department of Astronomy and Astrophysics, University of Chicago, Chicago, IL 60637}}

\maketitle

\begin{abstract}
GRB 970508 is the second gamma-ray burst (GRB) for which an optical afterglow 
has been detected.  It is the first GRB for which a distance scale has been 
determined:  absorption and emission features in spectra of the optical 
afterglow place GRB 970508 at a redshift of $z \ge 0.835$ [1,2].  The lack of a 
Lyman-$\alpha$ forest in these spectra further constrains this redshift to be 
less than $z \sim 2.1$.  I show that the spectrum of the optical afterglow of 
GRB 970508, once corrected for Galactic absorption, is inconsistent with the 
relativistic blast-wave model unless a second, redshifted source of extinction 
is introduced.  This second source of extinction may be the yet unobserved host 
galaxy.  I determine its redshift to be $z = 1.09^{+0.14}_{-0.41}$, which is 
consistent with the observed redshift of $z = 0.835$.  Redshifts greater than 
$z = 1.40$ are ruled out at the 3 $\sigma$ confidence level.
\end{abstract}

\section*{Introduction}
Discovered by the BeppoSAX Gamma-Ray Burst Monitor [3], GRB 970508 is the 
second gamma-ray burst (GRB) for which an optical afterglow has been detected 
(e.g., [4]).  Transient X-ray, near-infrared, millimeter, and radio emission 
have also been detected.  GRB 970508 is the first GRB for which a distance 
scale has been determined:  Metzger et al. [1,2] report the existence of 
absorption and emission features in spectra of the optical afterglow taken with 
the Keck II 10-m telescope $\approx$ 2 days and $\approx$ 26 days after the GRB 
event.  Their identification of these features places GRB 970508 at a redshift 
of $z \ge 0.835$.  The lack of a Lyman-$\alpha$ forest in these spectra further 
constrains this redshift to be less than $z \sim$ 2.1.  Consequently, GRB 
970508 is almost certainly cosmological in origin.
 
A host galaxy for GRB 970508 has not yet been observed.  However, Metzger et 
al. [1] report that the line emission observed in the Keck II spectra is 
consistent with constancy between their May 11 and June 5 observations.  Over 
this same period, the continuum emission has faded.  This suggests that a host 
galaxy of relatively weak continuum emission may be present at the observed 
redshift of $z = 0.835$ [2].
 
GRB afterglows are believed to be described by the relativistic blast-wave 
model [5-13].  Furthermore, the afterglow of GRB 970228, the only other GRB for 
which sufficient afterglow information is available, has been shown to be 
compatible with this model [14,11,15-17,13].  A basic prediction of the 
relativistic blast-wave model is that after an initial period of increasing 
optical flux, lasting hours to days, the optical flux of the afterglow will 
decrease as a power-law of index $-1.5 \lesssim b \lesssim -0.5$.  During this 
period of declining optical flux, the optical spectrum will be described by a 
power-law of index $a = 2b/3$ (e.g., [15]).  However, in the case of GRB 
970508, $a \approx b$ [18], which is inconsistent with the relativistic 
blast-wave model.

In \S2, I show that the reported GRB 970508 optical afterglow measurements, 
once corrected for Galactic absorption, are inconsistent with this prediction 
of the relativistic blast-wave model unless a second, redshifted source of 
extinction is introduced.  This second source of extinction may be the yet 
unobserved host galaxy.  I determine its redshift and I estimate its hydrogen 
column density along the line of sight.  An observing strategy for future 
optical afterglow observations is recommended in \S3.

\section*{Data Analysis \& Model Fit}
As of 1997 November 3, $>$ 70 optical and near-infrared measurements 
of the GRB 970508 afterglow have been reported.  These measurements span $>$ 80 
days, the earliest of which was taken $\approx$ 4 hours after the GRB event.  
Although the photometry is generally quite good, with quoted errors that are 
often less than 0.1 mag, zero-point errors are evident between different 
observing groups in various bands.  Consequently, I consider only the largest 
self-consistent subset of these data:  Sokolov et al. [18] report 22 optical 
measurements (B, V, R$_C$, and I$_C$ bands) taken with a 6-m telescope between 
$\approx$ 2 days and $\approx$ 31 days after the GRB event.  All of these 
measurements were taken after the optical flux had peaked (\S1), also $\approx$ 
2 days after the GRB event.  
 
I have corrected these measurements for Galactic absorption using the IRAS 100 
$\mu$m V-band absorption measure of Rowan-Robinson et al. [19] and the 
interstellar absorption curve [20,21].  The V-band correction is $A_V = 0.09$ 
mag.  I use the absorption measure of Rowan-Robinson et al., which measures the 
dust directly, instead of measures of the hydrogen column density, i.e. 
[22-24], since the IRAS 100 $\mu$m flux about the location of GRB 970508 varies 
significantly on angular scales that are smaller than the scales over which 
these measures of the hydrogen column density are averaged.  The corrected 
measurements are plotted in Figure 1.
\input psfig
\begin{figure}[b!] 
\centerline{\psfig{file=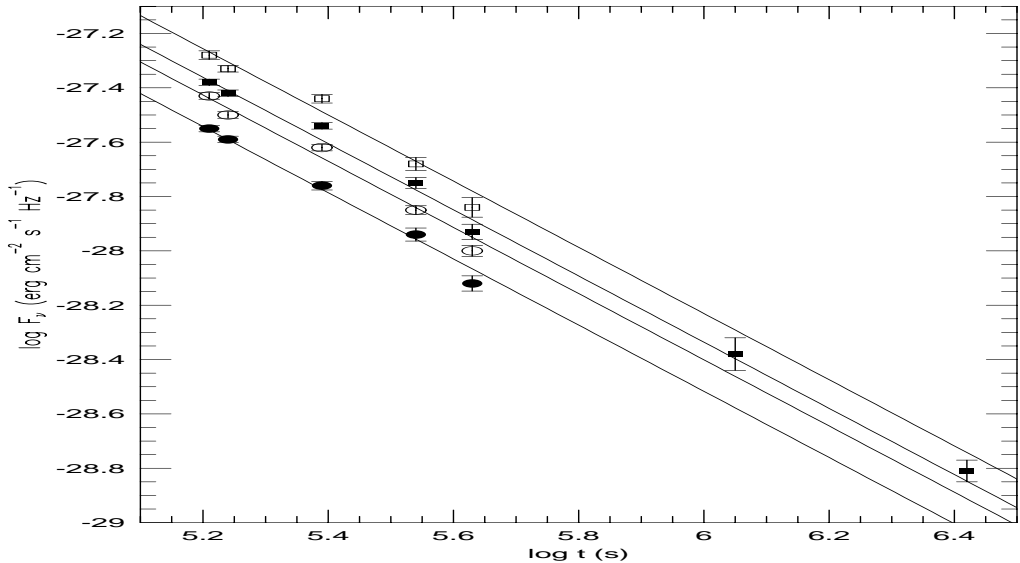,height=3.5in,width=6.5in}}
\caption{{Fluxes of Sokolov et al. [18] corrected for Galactic absorption (\S2) and the best fit to equation (1).  Solid circles are B-band measurements, open circles are V-band measurements, solid squares are R$_C$-band measurements, and open squares are I$_C$-band measurements.}}
\end{figure}

The following model is now $\chi^2$-fitted to these corrected measurements:
\begin{equation}
F_\nu = F_0\nu^{\frac{2b}{3}}t^b - F_{ext}(\nu;A_V(z),z).
\end{equation}
The first term is the extinction-free prediction of the relativistic blast-wave 
model (\S1).  The second term is the correction that a second, redshifted 
source of extinction introduces.  It is given by redshifting the interstellar 
absorption curve and by specifying the magnitude of the extinction, which I 
parameterize as the V-band absorption magnitude {\it at the redshift of the 
source}:  $A_V(z)$.  A constant error of $\approx$ 0.07 mag must be added in 
quadrature to the quoted errors of Sokolov et al. [18] for the model to fit the 
data ($\chi^2 \approx \nu$).  This suggests that either the quoted errors are 
underestimated by a factor $\sim 2$, or that the flux is varying by $\approx$ 
6\% on timescales of days.  Reichart [16] noticed possibly related temporal 
behavior in the light-curve of GRB 970228.
 
The best fit is:  $\log F_0 = -9.03^{+0.44}_{-0.44}$ cgs, $b = 
-1.22^{+0.03}_{-0.03}$, $A_V(z) = 0.24^{+0.12}_{-0.08}$ mag, and $z = 
1.09^{+0.14}_{-0.41}$.  The quoted uncertainties are 1 $\sigma$ confidence 
intervals for one interesting parameter.  The best-fit redshift is consistent 
with the observed redshift of $z = 0.835$.  Furthermore, redshifts greater than 
$z = 1.40$ are ruled out at the 3 $\sigma$ confidence level.  The best-fit 
V-band absorption magnitude corresponds to a hydrogen column density of 
$\approx$ 4.5 x 10$^{20}$ cm$^{-2}$.  The possibility that there is no second 
source of absorption, i.e., $A_V(z) = 0$, is ruled out at the 3.8 $\sigma$ 
confidence level.

\section*{Discussion \& Conclusions}
Using Equation (1) and the best-fit temporal decline power-law index, I have 
scaled each of the measurements that were fitted to in \S2 to its corresponding 
value for May 11, just shortly after the optical peak.  These points define the 
post-peak, time-independent optical spectrum and are plotted in Figure 2.  The 
best-fit spectrum, as well as the extinction-free, relativistic blast-wave 
component of this spectrum (the first term of equation (1)), are also plotted 
in Figure 2.  A broad absorption feature is apparent in the best-fit spectrum.  
This is the ultraviolet absorption feature of the interstellar absorption 
curve, redshifted into the B band.  Had the redshift of GRB 970508 been less 
than $z = 0.835$, U-band measurements would also have been necessary for this, 
the only strong feature of the interstellar absorption curve, to have been 
detected.  Consequently, in this type of analysis, different bands most 
sensitively probe different redshift ranges, depending on whether or not this 
absorption feature has been redshifted into the band in question.  Since GRBs 
are generally believed to have redshifts of $z \sim 1$, self-consistent sets of 
U-, B-, and V-band measurements should be a goal of future optical afterglow 
observations.
\begin{figure}[b!] 
\centerline{\psfig{file=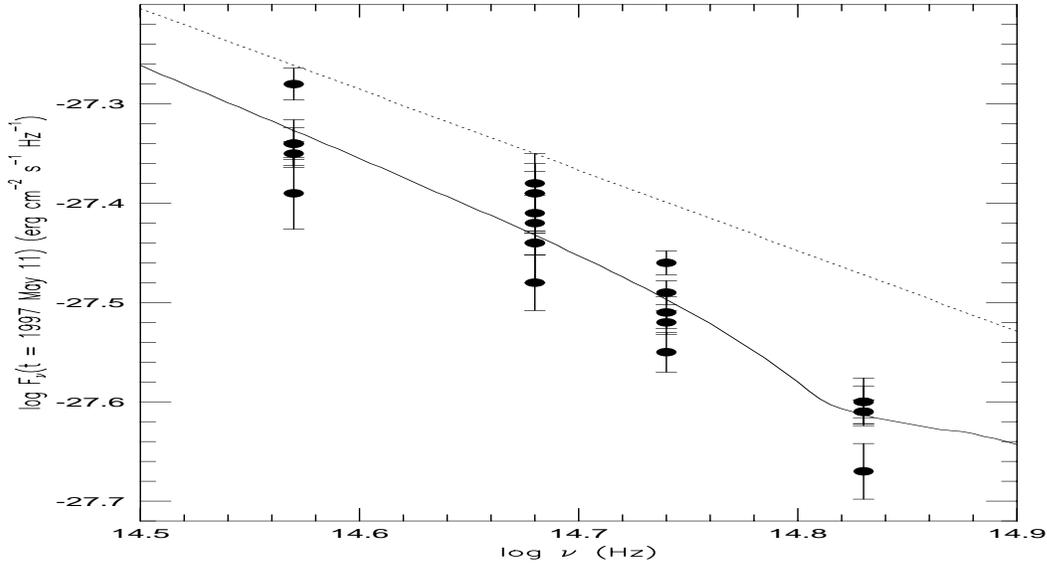,height=3.5in,width=6.5in}}
\caption{{The post-peak, time-independent optical spectrum of the afterglow of GRB 970508, scaled to 1997 May 11.  Plotted from left to right are the I$_C$-, R$_C$-, V-, and B-band fluxes of Sokolov et al. [18] corrected for Galactic absorption.  The solid line is the best-fit spectrum and the dotted line is the extinction-free, relativistic blast-wave component of this spectrum.}}
\end{figure}

In this letter, I present a method by which redshifts can be determined for 
GRBs that are associated with host galaxies, even if absorption or emission 
lines are not observable or if spectra of sufficient quality are unattainable.  
This method also yields hydrogen column densities along the line of site, which 
provides valuable information about the distribution of GRBs within galaxies.  
For GRB 970508, I find that the redshift of its host galaxy, or possibly that 
of an intermediate galaxy, is $z = 1.09^{+0.14}_{-0.41}$, which is consistent 
with the observed redshift of Metzger et al.:  $z = 0.835$.  Redshifts greater 
than $z = 1.40$ are ruled out at the 3 $\sigma$ confidence level.
 
\acknowledgments
I am grateful to D. Q. Lamb, R. C. Nichol, B. P. Holden, F. J. Castander, W. B. 
Burton, and D. Hartmann for valuable discussions and information.

\end{document}